\documentclass[acus]{JAC2000}

\usepackage{graphicx}

\begin{document}
\title{\flushright{T14}\\[15pt] \centering Luminosity Measurement at PEP-N 
}

\author{Mark Mandelkern, University of California, Irvine, CA92697, USA}

\maketitle

\begin{abstract}

The PEP-N experiment requires a fast on-line luminosity monitor of modest
accuracy plus an off-line method of determining integrated luminosity with
accuracy of 0.01 for each pb$^{-1}$. We propose the PEP-2 monitor, based
on observing single bremsstrahlung at zero degrees to the positron
direction at collision for the former and the use of Bhabha scatters at
polar angles $>$.03 radians for the latter requirement.

\end{abstract}

\section{On-Line Luminosity}

 An on-line monitor is required for tuning and monitoring the
machine. It is desirable that it provide a measurement with 10\% or better
accuracy, and fluctuations of less than 1\% at a refresh time of less than
1 second.  The PEP-2 monitor, based on observing single bremsstrahlung at
zero degrees to the positron direction at collision, described in Ref.
\cite{bib:field} seems appropriate.

 Single bremsstrahlung, or radiative Bhabha scattering, has a 
differential cross section, integrated over electron and positron angles, 
of:

\begin{equation}
\frac{d\sigma}{d\omega}=\frac{4\alpha r_0^2}{\omega} 
\frac{E-\omega}{\omega}
(V-2/3)[ln\frac{m}{q_{min}}-1/2]
\end{equation}
where $V=\frac{E-\omega}{E}+\frac{E}{E-\omega}$ and 
$q_min=\frac{m}{4\gamma^2}\frac{\omega}{E-\omega}$. Here $E$ is the 
initial electron or positron energy, $\gamma=E/m$ and $r_0=e^2/m$. The 
angular distribution of the $\gamma s$ is strongly forward with angular 
width $\sim\gamma^{-1}$. $\frac{d\sigma}{d\omega}$ is a function only of 
$\omega/E$ so the flux of $\gamma s$ at $\sim 0^\circ$ to the LER is 
independent of $s$. For PEP-N conditions I have used the program BBBREM 
\cite{bib:bbbrem}, provided by Lew Keller, to estimate the cross section 
for $\omega>400$ MeV radiation from the $e^+$ beam to be 76 mb.

 The momentum transfer for this process can be remarkably small,
corresponding to a very large impact parameter $\rho$ and leading to
screening effects which must be taken into account. If we choose E=3 GeV
and $\omega>300 MeV$, $q_{min}=0.4 10^{-9}$ MeV and $\rho_{max}=0.05 cm$
which is greater than the transverse size of the beams in PEP-N. The
consequence is that the cross section is cut off at a momentum transfer
$\sim q_{min}$. This problem has been treated by various authors and the
following result by Burov and Derbenev is quoted by Ref. \cite{bib:blinov}
for the case of a for a Gaussian beam density where the transverse beam
size is smaller than characteristic impact parameters:

\begin{equation}
\frac{d\sigma}{d\omega}=\frac{4\alpha 
r_0^2}{\omega}\frac{E-\omega}{\omega}
(V-2/3)[ln\frac{\Delta_y\Delta_z}{\lambda_C(\Delta_y+\Delta_z)}+
ln2+c/2+\frac{V-5/9}{V-2/3}] 
\end{equation}
 where $c=0.577$ and $\Delta_y$ and $\Delta_z$ are the rms transverse beam
dimensions. $\lambda_C$ is the electron Compton wavelength ($m^{-1}$).  
The sensitivity of this effective cross section to variation of the PEP-N
beam is approximately a 3.5\% increase for a doubling of the radius.  
Despite this modest sensitivity, the dependence on beam size and shape
introduces uncertainty that is undesireable for an absolute luminosity
measurement. The background to radiative Bhabhas at $0^{\circ}$ is
synchrotron radiation and beam-gas bremsstrahlung. At PEP-II, a Cerenkov
shower counter is used with a threshold sufficiently high to be immune to
the SR. The beam-gas background is apparently not a problem.\\

 The interaction region should be designed so that such a monitor
can be installed, which requires a clear aperture, suitable window, and
space for the monitor. At PEP-2, the monitor is installed at 8m from the
interaction point. We also want this monitor well downstream of the
detector.\\

 \section{Off-Line Luminosity}

 The accurate and precise determination of integrated luminosity
required for the experiment will be obtained from QED processes observed
in the detector. We require a 1\% or better measurement for each inverse
picobarn of running. The available processes are Bhabha scattering and
annihilations into muon pairs and gammas. We consider them individually in
the context of the standard detector design. Our luminosity determination
will be similar to that of BABAR, described for example in Touramanis'
talk at the 2/2001 BABAR Collaboration Meeting. The BABAR determination is
based on wide-angle ($>45^{\circ}$) Bhabhas and muon pairs. The systematic
error is contributed to by the Monte Carlo (1-2\%) and cut stability
(1\%), for an overall ~2\%. The annihilation to 2 photons has a greater
systematic uncertainty, at least 3\%, since the event rate is sensitive to
mass and the geometrical acceptance is less well defined (angles for
photons are not measured as well as those for charged particles).  \\

 In PEP-N the experimental situation is somewhat different. Since
the calorimeter has relatively course spatial resolution ($\sigma \sim
2.5$ cm), it is not possible to accurately define the acceptance for
photons, leading to an unacceptably large systematic error for the 2
photon annihilation rate.  Since the luminosity is much smaller than for
BABAR and we seek 1\% uncertainties on a point-by-point basis, we must
accept Bhabha and especially muon pair events at smaller polar angles,
which requires good angular measurements at small angles to adequately
define the acceptance. To obtain a 1\% statistical error for each inverse
pb we require $>10,000$ events for an integrated cross section of $>10$
nb. On the other hand the PEP-N detector is simpler and we may do better
in the Monte Carlo simulation, which is the dominant error for the BABAR
luminosity. In particular one particle for all Bhabha and muon pair events
will be seen by the forward planar tracking chamber and electromagnetic
and hadron calorimeters. \\

 \subsection{Geometry}

 These (approximate) geometrical parameters are taken from the
current detector layout. The beam pipe is expected to have a 5 cm radius
and the default is 2.5 mm of aluminum. We assume $4\pi$ tracking with 200
micron resolutiom for radii $<60$ cm , planar forward tracking with 200
micron resolution at z=120 cm with unhindered aperture of $\pm
23^{\circ}$, planar forward electromagnetic calorimetry at z=180 cm with
$\pm 36^{\circ}$ aperture and planar forward hadron calorimetry at z=220
cm with $\pm 27 ^{\circ}$ aperture.  The forward hadron calorimeter will
be used for muon ID. \\

 \subsection{Bhabhas}

\begin{figure*}[t]
\centering
\includegraphics*[width=150mm]{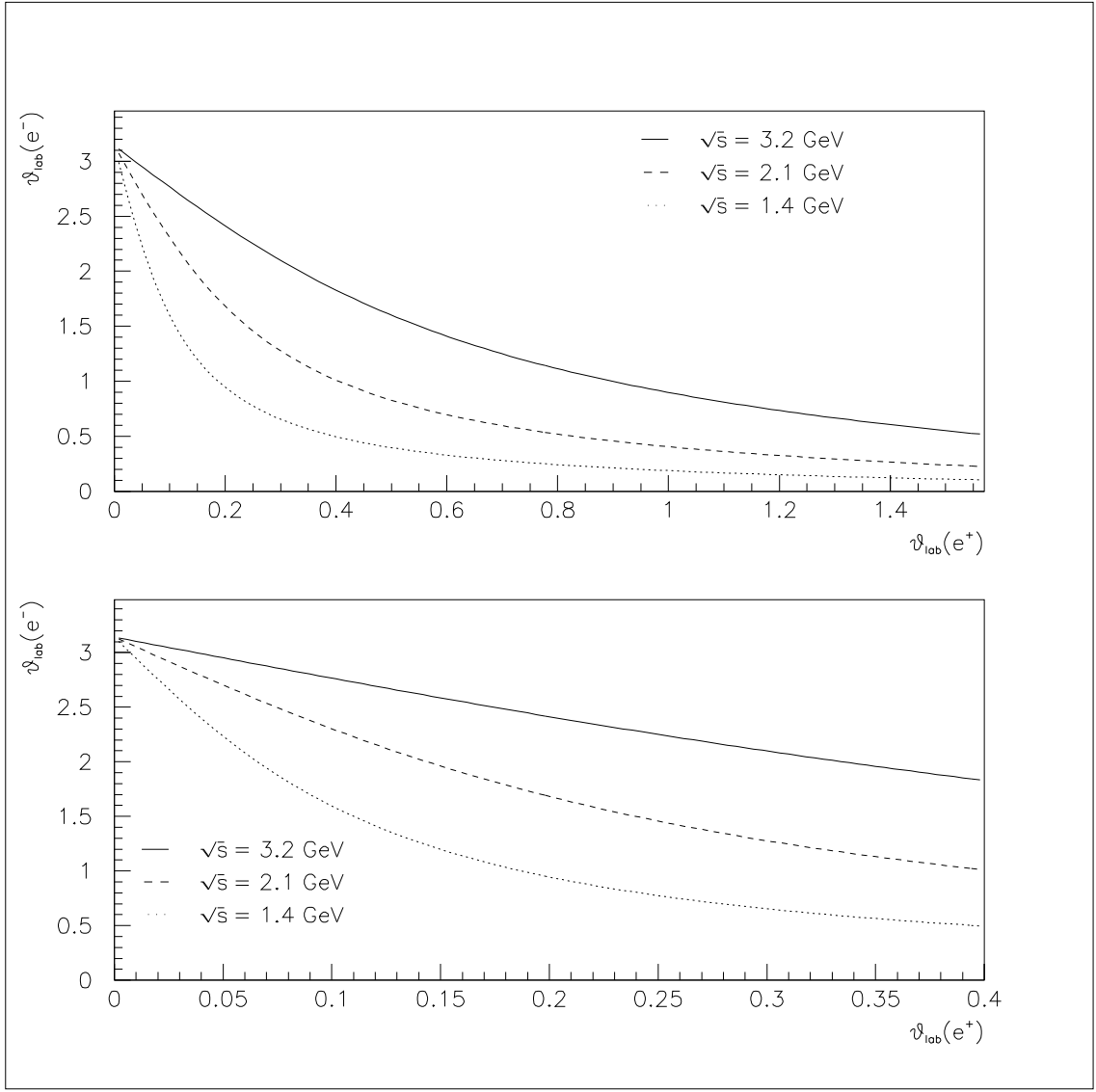}
\caption{Bhabhas: electron v. positron laboratory angles.}
\label{fig:p2}
\end{figure*}

\begin{figure*}[t]
\centering
\includegraphics*[width=150mm]{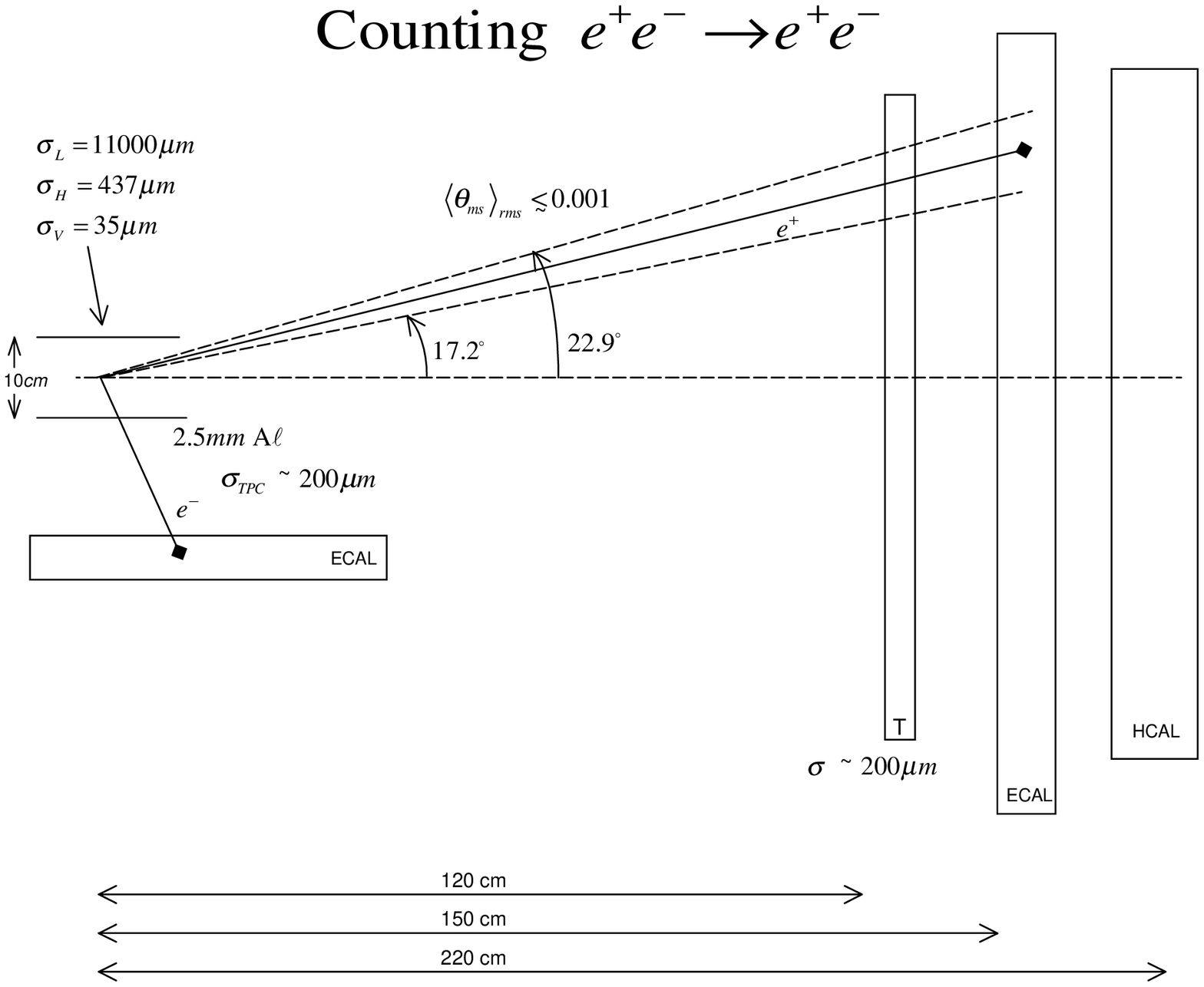}
\caption{Geometry for counting Bhabha scatters.}
\label{fig:geom}
\end{figure*}

 Both electron and positron can be identified at all angles since
we have nearly $4\pi$ tracking and electromagnetic calorimetry. In order
to get adequate statistics we must take advantage of the large forward
cross section and count events in which one particle strikes the forward
tracking chamber and forward electromagnetic calorimeter. It will
certainly be helpful to identify the backward electron as well. The cross
section, as seen in Table \ref{tab:bhabha} is well over 100 nb at all
energies.  For good control of systematics, it will be useful to define an
acceptance at a relatively large positron angle. This avoids relying on
events in which the e$^+$ passes very obliquely through the beam pipe and
reduces the angular accuracy and precision required to define the
acceptance. However we wish events in which the forward track passes
directly into the forward tracking chamber, missing the barrel
calorimeter, as shown for example in Fig. \ref{fig:geom}. We give cross
sections integrated between positron laboratory angles of 0.3
($17.2^{\circ}$) and 0.4 ($22.9^{\circ}$). As seen in Figure \ref{fig:p2},
the corresponding electron appears at $28^\circ$-$40^\circ$ at $\sqrt
s=1.4$ GeV and $97^\circ$-$114^\circ$ at $\sqrt s=3$ GeV, and is detected
in the barrel calorimeter which extends backward to $157^\circ$. We will
not be limited statistically in the Bhabha measurement. The acceptance
determination requires that we measure angles to about 1.5 mr which should
be relatively straightforward using the well defined interaction point and
the forward tracking chamber about 120 cm from the interaction point with
spatial resolution $\sim 200 \mu$m.  Multiple scattering is a
consideration here. At $17.2^{\circ}$, the effective thickness of the 2.5
mm Al beam pipe is .095 radiation lengths for a rms multiple scattering
angle of 1.1 mr. We can't tolerate a much thicker beam pipe.\\

 \subsection{Muon pairs}

 The muon pair cross section is much smaller and to obtain
adequate statistics we would have to accept events at much smaller angles.  
Table \ref{tab:muon} gives the integrated cross section between laboratory
angles of 0.1 ($5.7^{\circ}$) and 0.4 ($22.9^{\circ}$). Even so the
statistics will be marginal at the largest center of mass energies. The
smaller angles would then require more precise angular measurements for
the acceptance determination, i.e. about 0.5 mr. However the multiple
scattering for a very forward muon passing obliquely through the beam pipe
is much larger, i.e. at $5.7^{\circ}$, the effective beam pipe thickness
is about 28\% of a radiation length and the rms multiple scattering angle
is about 2 mr. A substantially thinner beam pipe would be required, or one
with an angled window which is not obviously feasible at small angles.  
Muon pairs will be useful as a rough check of the Bhabha measurement but
it will hard to obtain a precise luminosity because of statistical and
systematic uncertainties. \\

 \subsection{Conclusion}

 Using Bhabhas, the PEP-N detector as proposed should produce
integrated luminosity measurements with the desired 1-2\% accuracy for
individual points representing about 1 pb$^{-1}$ of integrated luminosity.
Muon pairs will be useful as a check although the muon pair luminosity
will not generally have the required statistical accuracy.

\newpage
\begin{table}
\begin{center}
\begin{tabular}{|c|c|c|c|c|c|c|c} 
\hline
$e^-$ energy & $E_{cm}$ & $\theta^l_{min}$ & $\theta^l_{max}$ & 
cos($\theta^{cm}_{max}$) & cos($\theta^{cm}_{min}$) & $\sigma$(nb)\\ 
\hline
\hline
0.100 &  1.114 &  0.300 &  0.400 &  0.171 & -0.120 & 280.499\\ \hline
0.200 &  1.575 &  0.300 &  0.400 &  0.477 &  0.222 & 174.436\\ \hline
0.300 &  1.929 &  0.300 &  0.400 &  0.618 &  0.404 &  152.080\\ \hline
0.400 &  2.227 &  0.300 &  0.400 &  0.699 &  0.517 &  143.612\\ \hline
0.500 &  2.490 &  0.300 &  0.400 &  0.752 &  0.594 &  139.480\\ \hline
0.600 &  2.728 &  0.300 &  0.400 &  0.789 &  0.650 &  137.151\\ \hline
0.700 &  2.946 &  0.300 &  0.400 &  0.816 &  0.692 &  135.706\\ \hline
0.800 &  3.150 &  0.300 &  0.400 &  0.837 &  0.725 &  134.748\\ \hline
0.900 &  3.341 &  0.300 &  0.400 &  0.854 &  0.752 &  134.080\\ \hline
1.000 &  3.521 &  0.300 &  0.400 &  0.868 &  0.774 &  133.596\\ \hline

\end{tabular}
\end{center}
\label{tab:bhabha}
\caption{Cross sections for Bhabhas.}
\end{table}

\begin{table}
\begin{center}
\begin{tabular}{|c|c|c|c|c|c|c|c}
\hline
$e^-$ energy & $E_{cm}$ & $\theta^l_{min}$ & $\theta^l_{max}$ &
cos($\theta^{cm}_{max}$) & cos($\theta^{cm}_{min}$) & $\sigma$(nb)\\ 
\hline
\hline
    0.100 &  1.114 &  0.100 &  0.400 &  0.856 & -0.120 & 62.416\\ \hline
    0.200 &  1.575 &  0.100 &  0.400 &  0.925 &  0.222 & 25.226\\ \hline
    0.300 &  1.929 &  0.100 &  0.400 &  0.950 &  0.404 & 14.103\\ \hline
    0.400 &  2.227 &  0.100 &  0.400 &  0.962 &  0.517 &  9.093\\ \hline
    0.500 &  2.490 &  0.100 &  0.400 &  0.969 &  0.594 &  6.372\\ \hline
    0.600 &  2.728 &  0.100 &  0.400 &  0.974 &  0.650 &  4.721\\ \hline
    0.700 &  2.946 &  0.100 &  0.400 &  0.978 &  0.692 &  3.641\\ \hline
    0.800 &  3.150 &  0.100 &  0.400 &  0.981 &  0.725 &  2.895\\ \hline
    0.900 &  3.341 &  0.100 &  0.400 &  0.983 &  0.752 &  2.358\\ \hline
    1.000 &  3.521 &  0.100 &  0.400 &  0.985 &  0.774 &  1.958\\ \hline

\end{tabular}
\end{center}
\label{tab:muon}
\caption{Cross sections for $\mu$ pairs.}
\end{table}

\begin{figure*}[t]
\centering
\includegraphics*[width=150mm]{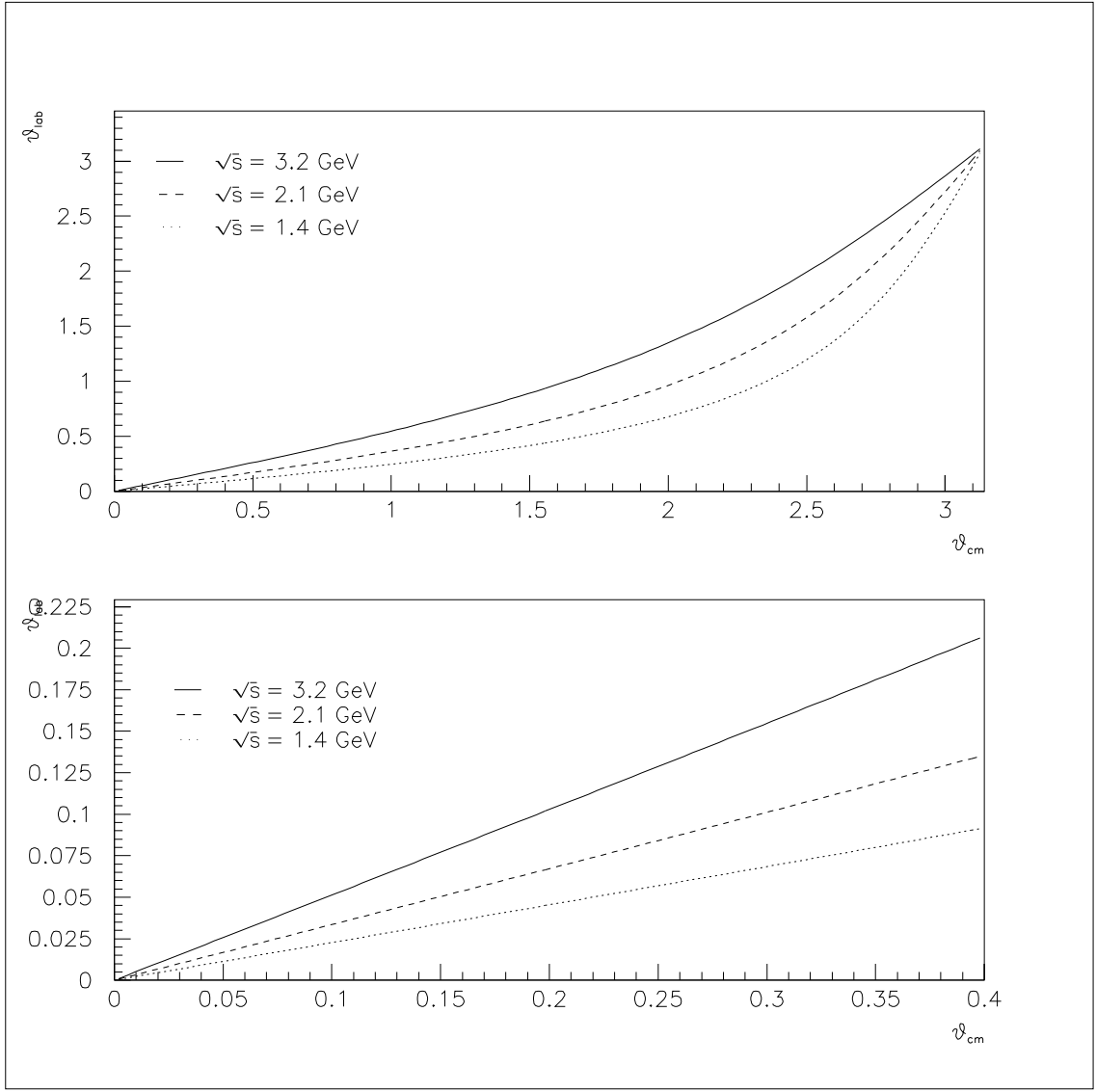}
\caption{Bhabhas:laboratory v. center of mass angles.}
\label{fig:p1}
\end{figure*}

\end{document}